\documentclass[amstex,aps,showpacs,preprint]{revtex4}%
\usepackage{graphicx}
\usepackage{color}
\usepackage{bm}
\usepackage{longtable}
\usepackage{amsmath}
\usepackage{amsfonts}
\usepackage{amssymb}%
\usepackage[normalem]{ulem}
\usepackage{soul}

\begin{document}
\title{Three-box paradox and ``Cheshire cat grin'': the case of spin-1 atoms}
\author{A. Matzkin}
\affiliation{Laboratoire de Physique Th\'{e}orique et Mod\'{e}lisation, CNRS Unit\'{e}
8089, Universit\'{e} de Cergy-Pontoise, 95302 Cergy-Pontoise cedex, France}
\author{A. K. Pan}
\affiliation{Graduate School of Information Science, Nagoya University, Chikusa-ku, Nagoya
464-8601, Japan}

\begin{abstract}
We propose in this work a definite theoretical implementation of the
three-box paradox -- a scheme in which a single quantum particle
appears to be present with certainty in two separate boxes -- with
spin-1 atoms. We further show how our setup can give rise to a
``Cheshire cat grin'' type of situation, in which an atom can
apparently be found with certainty in one of the boxes while one of
its properties (the angular momentum projection along a specifically
chosen axis) appears to be in a different box. The significance of
our findings are discussed relative to the status of the properties
of a system obtained from weak measurements.

\end{abstract}

\maketitle

\section{Introduction}

What is the value of a physical property prior to a measurement or
between two measurements? In the context of standard quantum
mechanics the question does not make sense: trying to answer the
question implies disturbing the system thereby changing the nature
of the experiment. Appealing to counterfactual arguments to
understand the situation rarely helps, as one is usually led to
apparent paradoxes and odd behavior, such as answering the ``Which
path?'' question in the well-known Wheeler delayed choice experiment
\cite{wheeler}.

The three box thought-experiment proposed by Aharonov and Vaidman
\cite{tbp} leads to that type of situation. The setup involves three
separate boxes $A,$ $B$ and $C$ and one particle. Given suitable
initial and final wavefunctions (consisting of a cleverly chosen
superposition of the particle being in one of the boxes), one wants
to know in which box the particle can be found at some intermediate
time. The quantum formalism seems to indicate that if a measurement
could be made to detect the particle in, say box $A$, while allowing
the system to reach the final state, then the particle would be
found in box $A$ with certainty. But if box $B$ was opened instead,
then the particle would also be found there with certainty, although
it is of course impossible to find with certainty a single particle
in two different boxes. A closely related setup involves a particle
and one of its property, e.g., the spin: the particle appears to have
taken with certainty one of two given paths (on the ground that if
the particle presence along the other path could be measured --
while allowing the system to reach the final state -- the
probability of finding it there would be zero) while the spin
appears to have taken the other path with certainty.\ This setup,
which has recently received increased interest
\cite{cheshir1,cheshir2,cheshir3,peculiar}, was given \cite{aharonov
rohrlich book} the suggestive name of ``Cheshire cat'' since the
grin (the property) appears to be separated from the cat (the
particle).

These apparent paradoxes are based on counterfactual inferences --
opening a box disturbs the system which does not reach the final
state -- so that, irrespective of whether trying to answer the
``which path?'' or ``which box?'' question is or not legitimate, the
effects cannot be observed. However counterfactual reasoning can to
some extent be bypassed by employing a scheme known as Weak
Measurement (WM) \cite{WM}. In contrast to the usual projective
measurements, a WM consists of a weak unitary interaction coupling
the system with a meter.\ The system, largely unperturbed,
reaches the final state. The meter wavefunction correlated with
that final state indicates the \emph{weak value} of the weakly
measured system observable. WM\ thus appear as a tool to open
non-invasively the boxes and assert what is happening inside. Indeed
WMs have been employed experimentally in dozens of works, essentially
with optical setups, confirming the theoretical predictions (though
the interpretation of WM remains a controversial topic). In
particular an optical version of the three-box experiment was
realized \cite{exp} in a modified Mach-Zehnder style interferometer.

In this work, we propose a theoretical implementation of the
three-box setup and of a ``Cheshire cat grin'' scheme employing spin-
1 particles. We will have more specifically in mind spin-1\ atoms,
that have been extensively manipulated in atomic interferometry
experiments \cite{atomexp}, including the use of Stern-Gerlach type
of devices that are the essential tool in the setups presented
below. One motivation is that while an optical three-box experiment
can be explained in classical terms (based on classical interference
effects), this is of course not the case of experiments performed
with massive particles. Moreover, the theoretical account employing
spin-1 atoms involves the explicit wavepacket dynamical evolution,
contrary to the original idealized three-box thought-experiment.
Employing a well-defined physical system dispels in our view many
ambiguities that have given rise to controversies (eg Refs.
\cite{classical3,ravon vaidman07,soko}) discussed in relation with
the ideal three-box paradox.

This paper is organized as follows. The original three-box
paradoxes, either based on counterfactual arguments or on WMs are recalled in Sec.\ \ref{3bp}. The implementation of
the thought-experiment with spin-1 particles is described in Sec.
\ref{3bspin}. A Cheshire cat grin type of scheme, based on the setup
described in Sec. \ref{3bspin}, is developed in Sec.\ \ref{catsec},
with the "grin" taken to be the spin projection on a chosen axis.
The results and their significance along with some remarks in view
of a possible experimental implementation with atoms of the proposed
schemes is given in Sec. \ref{disc}.

\section{The three-box thought experiment: Counterfactuals and Weak
Measurements\label{3bp}}

\subsection{The three-box example and counterfactuals}

The three-box paradox \cite{tbp} is usually presented in the context of
time-symmetric quantum mechanics (TSQM) \cite{abl} as illustrating a complete
description of a quantum system at any time given a fixed initial (known as
``pre-selected") state and a fixed final (termed as ``post-selected") state. Assume
we have a quantum system that can be in one of the three boxes $A,B$ or
$C$.\ The mutually orthogonal states $\left\vert A\right\rangle ,$ $\left\vert
B\right\rangle $ and $\left\vert C\right\rangle $ label the particle being in
one of the respective boxes. Let the system be initially $(t=t_{i})$ prepared
in the state%
\begin{equation}
|\psi_{i}\rangle=\frac{1}{\sqrt{3}}\left(  |A\rangle+|B\rangle+|C\rangle
\right)  \label{1}%
\end{equation}
and post-selected at $t=t_{f}$ to the final state%
\begin{equation}
|\psi_{f}\rangle=\frac{1}{\sqrt{3}}\left(  |A\rangle+|B\rangle-|C\rangle
\right)  .\label{3}%
\end{equation}
What would happen if one of the boxes is opened at some intermediate
time between $t_{i}$ and $t_{f}$? Assume box $A$ is opened; if the
particle is found there, this means the initial state has been
projected to state $\left\vert A\right\rangle $ through
$\Pi_{A}\left\vert \psi_{i}\right\rangle $ where
$\Pi_{A}\equiv\left\vert A\right\rangle \left\langle A\right\vert .$
If the particle is not found there, then the state after box $A$ is
opened is $\Pi_{\bar{A}}\left\vert \psi_{i}\right\rangle$ where
\begin{equation}
\Pi_{\bar{A}}\equiv1-\Pi_{A}=\left\vert B\right\rangle \left\langle
B\right\vert +\left\vert C\right\rangle \left\langle C\right\vert \label{4}%
\end{equation}
is the complement of $\Pi_{A}$. However the transition amplitude
$\left\langle \psi_{f}\right\vert \Pi_{\bar{A}}\left\vert
\psi_{i}\right\rangle $ to the final post-selected state vanishes,
so the probability of not finding the particle in box $A,$
proportional to $\left\vert \left\langle \psi _{f}\right\vert
\Pi_{\bar{A}}\left\vert \psi_{i}\right\rangle \right\vert ^{2},$ is
zero: the conclusion is therefore that the particle \emph{must} have
been with certainty in box $A.$

A contradiction arises by repeating the same argument assuming now that box
$B$ is opened. If the particle is not found in box $B$, then the state after
box $B$ has been opened is $\Pi_{\bar{B}}\left\vert \psi_{i}\right\rangle $
with $\Pi_{\bar{B}}\equiv1-\Pi_{B}=\left\vert A\right\rangle \left\langle
A\right\vert +\left\vert C\right\rangle \left\langle C\right\vert $.\ But
$\left\vert \left\langle \psi_{f}\right\vert \Pi_{\bar{B}}\left\vert \psi
_{i}\right\rangle \right\vert ^{2}=0,$ so on its way from $\left\vert \psi
_{i}\right\rangle $ to $\left\vert \psi_{f}\right\rangle $ the particle
\emph{must} have been with certainty in box $B$! Of course, a single particle
cannot be with certainty in two different boxes at the same time, hence the paradox.

The paradox is apparently dissolved by remarking that quantum
mechanics does not allow this type of counterfactual reasoning.\ It
does not make sense, according to the standard interpretation, to
demand a ``complete'' description of the behavior of the system
between $\left\vert \psi_{i}\right\rangle $ and $\left\vert
\psi_{f}\right\rangle $ without actually making a measurement at box
$A$ and/or box $B$ that will disturb the system.\ In particular,
opening the two boxes jointly will yield a single particle either in
box $A$ or box $B$ (or none), and the quantum formalism consistently
predicts $\Pi_{A}\Pi _{B}\left\vert \psi_{i}\right\rangle =0.$
Nevertheless actually performing these measurements changes the
nature of the experiment: the particle never reaches the final state
$\left\vert \psi_{f}\right\rangle $, either because it is detected
in the box or because post-selection is not succesful if it is
undetected, leaving the original question relative to the behavior
of the system at an intermediate time unanswered.

\subsection{Weak measurements\label{swm}}

WM represents a tool that provides a certain type
of answer to the question. Briefly put, the main idea consists in
two steps: first a system in a preselected state  weakly
interacts unitarily with an apparatus, resulting in an entangled system-apparatus state; the
interaction couples a system observable $O$ with a dynamical
variable of the apparatus.\ Then a standard projective measurement
of a different system observable is made; one retains only the
outcomes leaving the system in the chosen post-selected state. The
corresponding projection leaves the apparatus
wavefunction in a certain final state.\ Under certain conditions,
basically amounting to a very weak interaction and widely
overlapping meter states \cite{sudarshan,pan matzkin12}, the final
state is simply shifted relative to the initial state, the shift
being proportional to the weak value $\left\langle O\right\rangle
_{w}$ of the system observable that was weakly coupled
to the meter.\ The weak value is given by%
\begin{equation}
\left\langle O\right\rangle _{w}=\frac{\left\langle \psi_{f}\right\vert
O\left\vert \psi_{i}\right\rangle }{\left\langle \psi_{f}\right\vert \left.
\psi_{i}\right\rangle }.\label{8}%
\end{equation}

In the present context,we see that WM allow to obtain information on some
system observable $O$ at some intermediate time while the system evolves from
$\left\vert \psi_{i}\right\rangle $ to $\left\vert \psi_{f}\right\rangle
.\ $Indeed, the weak coupling barely affects the system while the meter
wavefunction picks up a phase shift that can in principle be experimentally
detected.\ WM thus appears as a way to bypass counterfactual reasoning and
access to what is happening in a system between the initial and final states.

\subsection{The three box paradox}

The analysis of the three-box paradox with WM \cite{tbp,tollaksen07,ravon
vaidman07} involves replacing the measurements (box openings) and their
associated projectors $\Pi_{A},\Pi_{B},...$ with weak measurements and the
respective weak values $\left\langle \Pi_{A}\right\rangle _{w},$ $\left\langle
\Pi_{B}\right\rangle _{w},...$\ From the state definitions (\ref{1}) and
(\ref{3}) it is straightforward to apply Eq. (\ref{8}). This gives
\begin{equation}
\left\langle \Pi_{A}\right\rangle _{w}=1\text{ and }\left\langle \Pi_{\bar{A}%
}\right\rangle _{w}=0\label{10}%
\end{equation}
meaning that an apparatus weakly interacting with box $A$ will move, but that a
single meter that would weakly open boxes $B$ and $C$ jointly (see below for a
definite example in the context of spin-1\ particles) will not display any
shift. If WMs were to be interpreted along the same line as
projective measurements, the conclusion would be that on its way to
$\left\vert \psi_{f}\right\rangle $ the particle went through box $A$.\

However we also have
\begin{equation}
\left\langle \Pi_{B}\right\rangle _{w}=1\text{{ and }}\left\langle
\Pi_{\bar
{B}}\right\rangle _{w}=0\label{12}%
\end{equation}
leading to the conclusion that on its way to $\left\vert
\psi_{f}\right\rangle $ the particle went through box $B.$ But now,
unlike the case with projective measurements discussed above, no
counterfactual arguments are involved: if two  apparati open\
weakly boxes $A$ and $B$ respectively, we will have jointly
$\left\langle \Pi_{A}\right\rangle _{w}=1$ \emph{and} $\left\langle
\Pi _{B}\right\rangle _{w}=1$. In one sense the paradox is back
again, though whether with WM\ there is a real paradox involved or
whether WM allows to observe the superpositions typical of quantum
phenomena is largely a matter of interpretation (see Sec. \ref{disc}
below).
\begin{figure}[tb]
\includegraphics[angle=-90,width=13cm]{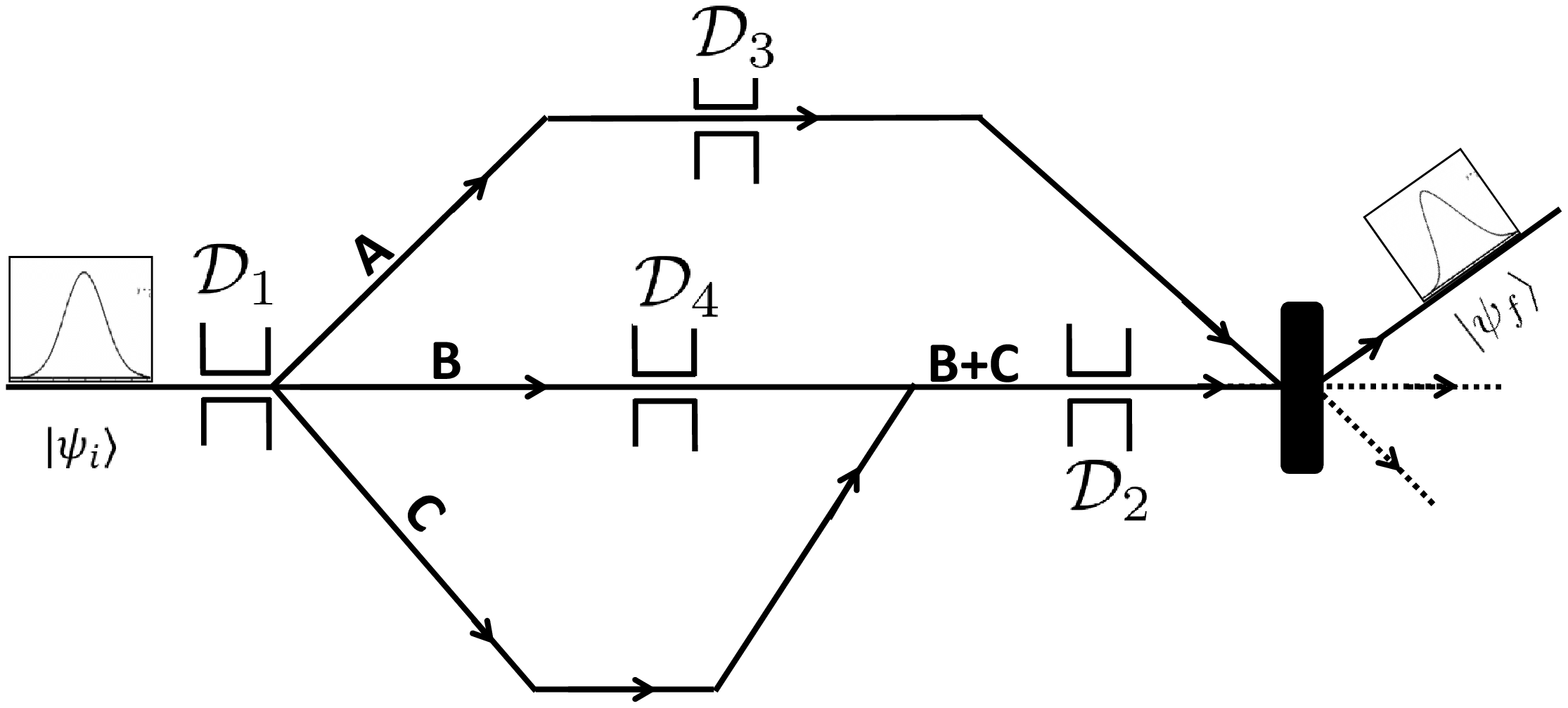}\caption{ The ``three box
paradox'' setup for spin-1 particles. Instead of three boxes, a particle prepared in an
initial pre-selected state $J=1$, $M=m_{i}$, enters a Stern-Gerlach type of
device $\mathcal{D}_{1}$ that separates the $m_{i}$ state on the $m_{\alpha}$
basis: the wavepacket is then divided along the three paths $A,B$ and $C$. The
wavepackets along $B$ and $C $ are recombined first, and then recombined with
the wavepacket traveling along path $A$, at which point a projective
measurement (represented by the black box) is made. The ``apparati''
$\mathcal{D}_{2},\mathcal{D}_{3}$ and $\mathcal{D}_{4}$ interact unitarily
with the system. }%
\label{fig j=1 setup}%
\end{figure}

\section{The three box paradox with spin-1 particles\label{3bspin}}

\subsection{General Remarks}

The three-box paradox, as initially proposed in \cite{tbp}, was
realized experimentally with photons \cite{exp}.\ As remarked by the
authors of Ref. \cite{exp}, the effects giving rise to the paradox
have a classical optical explanation in terms of the classical
interference of light. On the other hand spin-1 particles would
allow to realize experimentally the three-box paradox with matter
waves. Here we will simply describe the spin-1\ version of the
three-box paradox without consideration of any experimental
realization (a possible experimental realization is outlined in Sec.
\ref{disc}).

Assume spin-1 particles (e.g, atoms) are prepared in the initial state%
\begin{equation}
\left\vert \psi_{i}\right\rangle =\left\vert J=1,m_{z}=0\right\rangle
\left\vert \xi\right\rangle \label{ini}%
\end{equation}
where $\xi(\mathbf{r})\equiv \left\langle \mathbf{r} \right\vert
\left.  \xi \right\rangle$ is the spatial part of the wavefunction
and $\left\vert J=1,m_{z}=0\right\rangle \equiv\left\vert
m_{i}\right\rangle $ is the $J=1$ spin state (we will omit
explicitly denoting $J=1$ in the rest of the paper) with the spin
projection quantized along the $\mathbf{\hat{z}}$ axis with
azimuthal number $m_{z}=0$. We assume $\xi(\mathbf{r})$ can be
represented by a Gaussian, moving to the right on the
$\mathbf{\hat{y}}$ axis (see Fig. \ref{fig j=1 setup}), and its width
depending on the coherence length of the atoms; we will explicitly
write the spatial part of the wavefunction only when appropriate,
focusing on the sole spin part in most of the paper.

Assume that at $t=0$ the wavepacket enters a SG type of device ($\mathcal{D}%
_{1}$ in Fig. \ref{fig j=1 setup}) with an inhomogeneous magnetic field
directed along the direction $\mathbf{\hat{\alpha}}$. The effect of
$\mathcal{D}_{1}$ is to separate the wavepackets according to their associated
spin projection along $\mathbf{\hat{\alpha}}$. The spin part of the initial
state in the $\left\vert m_{\alpha}\right\rangle $ basis is transformed as%
\begin{equation}
\left\vert \psi_{i}\right\rangle =\left\vert m_{i}\right\rangle \left\vert
\xi\right\rangle \equiv\sum_{k=-1}^{1}\left\langle m_{\alpha}=k\right\vert
\left.  m_{z}=0\right\rangle \left\vert m_{\alpha}=k\right\rangle \left\vert
\xi(t=0)\right\rangle ,
\end{equation}
where
\begin{equation}
\left\langle m_{\alpha}\right\vert \left.  m_{\beta}\right\rangle \equiv
d_{m_{\alpha},m_{z}}^{J=1}(\beta-\alpha)
\end{equation}
is given by the reduced Wigner rotational matrix element. For $t>0$,
$\left\vert \xi\right\rangle $ separates into three wavepackets\ each
associated with a given value of $m_{\alpha}$, so that upon exiting
$\mathcal{D}_{1}$ the system wavefunction becomes%
\begin{equation}
\left\vert \psi(t)\right\rangle =\sum_{k}\left\langle m_{\alpha}=k\right\vert
\left.  m_{z}=0\right\rangle \left\vert m_{\alpha}=k\right\rangle \left\vert
\xi_{k}(t)\right\rangle ;\label{27}%
\end{equation}
the $\left\vert \xi_{k}\right\rangle $ can be computed by solving
the Schr\"{o}dinger equation inside $\mathcal{D}_{1}$ \cite{pan
matzkin12}. Each of the three paths is taken to represent a box: box
A is taken to be the $k=+1$ path, boxes B and C corresponding
respectively to the paths $k=0$ and $k=-1$. We assume that paths B
and C are recombined first and then recombined with path A (as shown
in the figure 1). These recombinations are assumed to take place
without affecting the spin state nor the phase difference. Finally a
projective measurement of the spin projection along the direction
$\mathbf{\hat{\phi}}$ is made at time $t_{f}$. The final
post-selected state is chosen to be
\begin{equation}
\left\vert \psi_{f}\right\rangle =\left\vert m_{f}\right\rangle \left\vert
\xi(t_{f})\right\rangle \equiv\sum_{k=-1}^{1}\left\langle m_{\alpha
}=k\right\vert \left.  m_{\phi}=+1\right\rangle \left\vert m_{\alpha
}=k\right\rangle \left\vert \xi(t_{f})\right\rangle \label{30}%
\end{equation}
with $\left\vert m_{f}\right\rangle \equiv\left\vert m_{\phi}=+1\right\rangle
.$

\subsection{Condition on path $A$\label{condA}}

In order to obtain the analogue of the three-box paradox, some transition
amplitudes must interfere destructively. Assume an apparatus $\mathcal{D}_{2}$
is positioned at $\mathbf{r}_{2}$ as indicated in Fig. \ref{fig j=1 setup}.
$\mathcal{D}_{2}$ measures the weak value of the projector $\Pi_{\bar{A}}$
[Eq. (\ref{4})] along the recombined path $B+C$. The projector $\Pi_{\bar{A}}$
can be taken here to project to a Gaussian centered on $\mathcal{D}_{2}$ whose
width encompasses the spatial extent of the spatial part of the wavepackets,
$\Pi_{\bar{A}}=\left\vert \Gamma\right\rangle \left\langle \Gamma\right\vert $
where%
\begin{equation}
\Gamma(\mathbf{r})=(\frac {2}{\pi\Delta^{2}})^{1/2}e^{-\left(  \mathbf{r}-\mathbf{r}_2%
\right)  ^{2}/\Delta^{2}}
%\frac{1}{{}}e^{\frac{-(\mathbf{r-r}_{2})^{2}}{2\Delta
%_{2}^{2}}}.
\end{equation}
The weak value $\left\langle \Pi_{\bar{A}}\right\rangle _{w}$ is
obtained from Eq. (\ref{8}) keeping in mind we are dealing with a
``real'' system endowed with dynamics (an element that is not taken
into account in the ideal three-box paradox presented in Sec.
\ref{3bp}). Denoting by $t_{2}$ the time at which $\mathcal{D}_{2}$
is triggered (and the weak measurement made) and introducing the
evolution operator $U(t_{k},t_{j})$ of the spin-1\ particle between
$t_{j}$ and $t_{k}$ we have
\begin{equation}
\left\langle \Pi_{\bar{A}}\right\rangle _{w}=\frac{\left\langle \psi_{f}%
(t_{f})\right\vert U(t_{f},t_{2})\Pi_{\bar{A}}U(t_{2},t_{i})\left\vert
\psi_{i}\right\rangle }{\left\langle \psi_{f}\right\vert U(t_{f},t_{2}%
)U(t_{2},t_{i})\left\vert \psi_{i}\right\rangle }.
\end{equation}
Following Eq. (\ref{27}), $U(t_{2},t_{i})\left\vert \psi_{i}\right\rangle $ is
of the form%
\begin{equation}
U(t_{2},t_{i})\left\vert \psi_{i}\right\rangle =\left\langle m_{\alpha
}=+1\right\vert \left.  m_{i}\right\rangle \left\vert m_{\alpha}%
=+1\right\rangle \left\vert \xi_{A}(t_{2})\right\rangle +\sum_{k=-1,0}%
\left\langle m_{\alpha}=k\right\vert \left.  m_{i}\right\rangle \left\vert
m_{\alpha}=k\right\rangle \left\vert \xi_{B+C}(t_{2})\right\rangle \label{13}%
\end{equation}
and $\Pi_{\bar{A}}\left\vert \xi_{A}(t_{2})\right\rangle $ vanishes (since
there is no spatial overlap between $\left\vert \Gamma\right\rangle $ and
$\left\vert \xi_{A}(t_{2})\right\rangle $). The weak value becomes%
\begin{equation}
\left\langle \Pi_{\bar{A}}\right\rangle _{w}=\frac{\left\langle \xi
(t_{f})\right\vert U(t_{f},t_{2})\Pi_{\bar{A}}\left\vert \xi_{B+C}%
(t_{2})\right\rangle }{\left\langle \psi_{f}\right\vert U(t_{f},t_{2}%
)U(t_{2},t_{i})\left\vert \psi_{i}\right\rangle }\left[  \sum_{k=-1,0}%
\left\langle m_{\alpha}=k\right\vert \left.  m_{i}\right\rangle \left\langle
m_{f}\right\vert \left.  m_{\alpha}=k\right\rangle \right]  \label{14}%
\end{equation}
Hence $\mathbf{\hat{\phi}}$ needs to be chosen (restricting
$\mathbf{\hat {\phi}}$ to lie in $yz$ plane) such that the
transition amplitudes $m_{\alpha}=0\rightarrow m_{\phi}=+1$ and
$m_{\alpha}=-1\rightarrow m_{\phi}=+1$ interfere destructively, viz.
by
solving%
\begin{equation}
\sum_{k=-1,0}\left\langle m_{f}\right\vert \left.  m_{\alpha}=k\right\rangle
\left\langle m_{\alpha}=k\right\vert \left.  m_{i}\right\rangle
=0\label{condition1}%
\end{equation}
One class of solutions to Eq. (\ref{condition1}) for a fixed value of $\alpha$
is given by
\begin{equation}
\phi=4\arctan\left(  \frac{8\tan^{3}\frac{\alpha}{4}-\frac{(3\cos
(2\alpha)+5)^{1/2}\sec^{6}\left(  \frac{\alpha}{4}\right)  }{2\sqrt{2}}%
}{\left(  \tan^{2}\frac{\alpha}{4}-1\right)  ^{3}}\right)  +4\pi
n.\label{fisol}%
\end{equation}
Therefore provided $\alpha$ and $\phi$ obey Eq. (\ref{fisol}), we
have $\left\langle \Pi_{\bar{A}}\right\rangle _{w}=0:$ the
apparatus $\mathcal{D}_{2}$ will not display
any change following post-selection.

Conversely if an apparatus $\mathcal{D}_{3}$ is positioned at $\mathbf{r}_{3}
$ along path $A$ as indicated in Fig. \ref{fig j=1 setup} the weak value of
the projector $\Pi_{A}$ is given by
\begin{equation}
\left\langle \Pi_{A}\right\rangle _{w}=\frac{\left\langle \psi_{f}%
(t_{f})\right\vert U(t_{f},t_{3})\Pi_{A}U(t_{3},0)\left\vert \psi
_{i}\right\rangle }{\left\langle \psi_{f}\right\vert U(t_{f},t_{3}%
)U(t_{3},t_{i})\left\vert \psi_{i}\right\rangle },
\end{equation}
where $t_{3}$ is the time at which the weak measurement is made. Employing
Eqs. (\ref{27}) and (\ref{30}) and keeping in mind $\Pi_{A}\left\vert \xi
_{k}(t_{3})\right\rangle =0$ for $k=B,C$ leads to%
\begin{equation}
\left\langle \Pi_{A}\right\rangle _{w}=\frac{\left\langle \xi_{f}%
(t_{f})\right\vert U(t_{f},t_{3})\Pi_{A}\left\vert \xi_{A}(t_{3}\right\rangle
\left\langle m_{f}\right\vert \left.  m_{\alpha}=1\right\rangle \left\langle
m_{\alpha}=1\right\vert \left.  m_{i}\right\rangle }{\sum_{k=-1}%
^{1}\left\langle m_{f}\right\vert \left.  m_{\alpha}=k\right\rangle
\left\langle m_{\alpha}=k\right\vert \left.  m_{i}\right\rangle }%
\end{equation}
which simplifies given the condition (\ref{condition1}) to%
\begin{equation}
\left\langle \Pi_{A}\right\rangle _{w}=\left\langle \xi(t_{f})\right\vert
U(t_{f},t_{3})\Pi_{A}\left\vert \xi_{A}(t_{3})\right\rangle .
\end{equation}
Hence the meter corresponding to the apparatus $\mathcal{D}_{3}$, interacting with the spin-1 system,
moves, the motion of the pointer being proportional to the system and meter
wavepackets overlap.\ Note that in the ideal case in which the projector
$\Pi_{A}$ perfectly overlaps with the system wavepacket at the time of
measurement, i.e. $\Pi_{A}\equiv\left\vert \xi_{A}(t_{3})\right\rangle
\left\langle \xi_{A}(t_{3})\right\vert $ we have (since $U(t_{f}%
,t_{3})\left\vert \xi_{A}(t_{3})\right\rangle =\left\vert \xi(t_{f}%
)\right\rangle )$%
\begin{equation}
\left\langle \Pi_{A}\right\rangle _{w}=1\label{18}%
\end{equation}
as in Eq. (\ref{10}) of the ideal three-box paradox.

\subsection{Condition on path $B$}

By employing the same reasoning followed for path $A,$ let us now position an apparatus $\mathcal{D}_{4}$ at $\mathbf{r}_{4}$ along path $B$ as indicated
in Fig. \ref{fig j=1 setup}. The weak value of the projector $\Pi_{B},$
measured by\ $\mathcal{D}_{4}$ at time $t_{4}$ is given by%
\begin{equation}
\left\langle \Pi_{B}\right\rangle _{w}=\frac{\left\langle \psi_{f}%
(t_{f})\right\vert U(t_{f},t_{4})\Pi_{B}U(t_{4},0)\left\vert \psi
_{i}\right\rangle }{\left\langle \psi_{f}\right\vert U(t_{f},t_{4}%
)U(t_{4},t_{i})\left\vert \psi_{i}\right\rangle }%
\end{equation}
taking the form%
\begin{equation}
\left\langle \Pi_{B}\right\rangle _{w}=\frac{\left\langle \xi_{f}%
(t_{f})\right\vert U(t_{f},t_{4})\Pi_{B}\left\vert \xi_{B}(t_{4}\right\rangle
\left\langle m_{f}\right\vert \left.  m_{\alpha}=0\right\rangle \left\langle
m_{\alpha}=0\right\vert \left.  m_{i}\right\rangle }{\sum_{k=-1}%
^{1}\left\langle m_{f}\right\vert \left.  m_{\alpha}=k\right\rangle
\left\langle m_{\alpha}=k\right\vert \left.  m_{i}\right\rangle }.
\end{equation}
By imposing the condition%
\begin{equation}
\sum_{k\neq0}\left\langle m_{f}\right\vert \left.  m_{\alpha}=k\right\rangle
\left\langle m_{\alpha}=k\right\vert \left.  m_{i}\right\rangle
=0\label{condition2}%
\end{equation}
we have%
\begin{equation}
\left\langle \Pi_{B}\right\rangle _{w}=\left\langle \xi_{f}(t_{f})\right\vert
U(t_{f},t_{4})\Pi_{B}\left\vert \xi_{B}(t_{4}\right\rangle =1\label{19}%
\end{equation}
(where the last equality is obtained only with an ideal projector $\Pi
_{B}=\left\vert \xi_{B}(t_{4})\right\rangle \left\langle \xi_{B}%
(t_{4})\right\vert $) and also%
\begin{equation}
\left\langle \Pi_{\bar{B}}\right\rangle _{w}=0.
\end{equation}
The result $\left\langle \Pi_{\bar{B}}\right\rangle _{w}=0$ can in
principle be checked by recombining the paths $A$ and $C$ and
performing a weak measurement along that recombined path, in full
analogy with the weak measurement of $\left\langle
\Pi_{\bar{A}}\right\rangle _{w}$.

The conditions to get vanishing transition elements (\ref{condition1}) and
(\ref{condition2}) can be solved jointly.\ The solution (for angles $\alpha$
and $\phi$ coplanar with the $z$ axis) is%
\begin{align}
\alpha &  =2\arccos\left(  \frac{1}{2}+\frac{\sqrt{5}}{10}\right)
^{1/2}\simeq63.4^{\circ}\\
\phi &  =2\arccos(\frac{1}{2}-\frac{1}{\sqrt{5}})^{1/2}\simeq153.4^{\circ}.
\end{align}
Analog solutions are readily obtained for other combinations (e.g, different
values of $m_{i},$ $m_{f}$) of pre-selected and post-selected states.

\begin{figure}[tb]
\includegraphics[angle=-90,width=13cm]{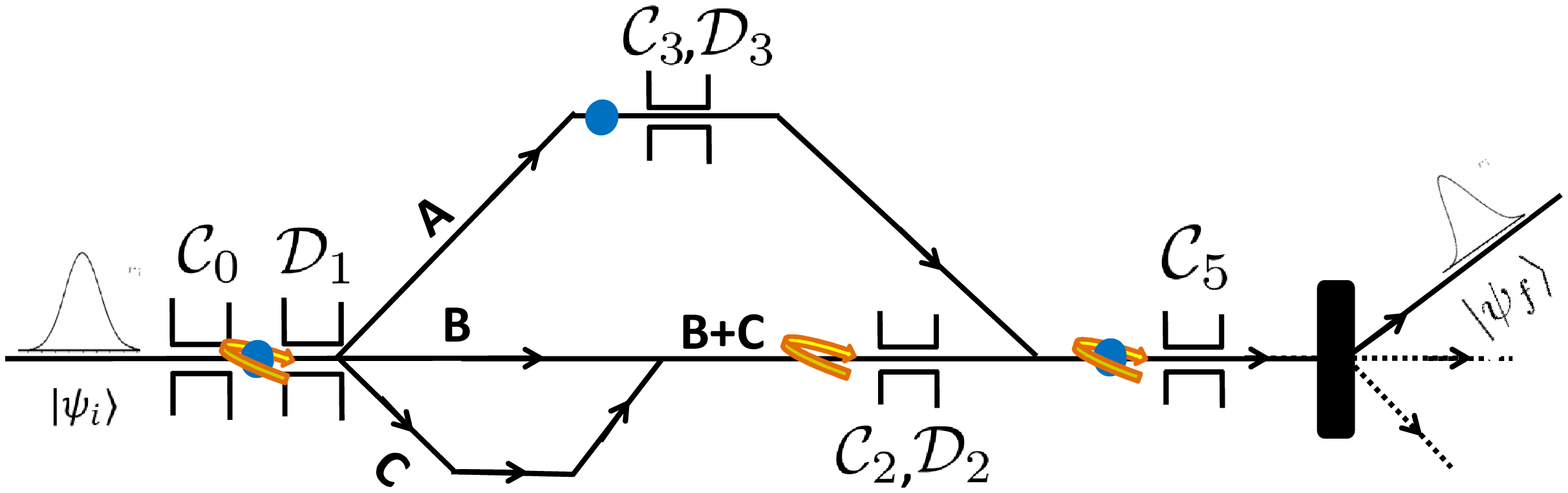}\caption{ The ``Cheshire cat''
setup for spin-1 particles. The particle (pictured by a blue circle) and the
spin projection $J_{\gamma}$ (represented by a rotating arrow) along a
specifically chosen direction $\mathbf{\hat{\gamma}}$ (defined in the text)
appear to travel along different paths. Indeed given the pre- and post-
selected states $|\psi_{i}\rangle$ and $|\psi_{f}\rangle$, an apparatus
positioned along $A$ detects the particle but no spin $J_{\gamma}$ whereas an apparatus
 along $B+C$ detects no particle but does detect its spin (see Sec.
\ref{catsec} for details).}%
\label{fig cheshire}%
\end{figure}

\section{The Cheshire cat grin with spin-1\ particles\label{catsec}}

The  ``grin without a cat'' that Alice experienced in Wonderland was
introduced in the context of weak measurement by Aharonov and
Rohrlich \cite{aharonov rohrlich book}. The idea is that given pre
and postselected states, the cat -- the particle -- can only be
found (by performing a WM) along a given box or path, whereas the
grin -- a property of the particle -- can only be found (by
performing another WM) along a different path. The idea has been
receiving increased interest recently both in refining theoretical
aspects and in the form of concrete proposals
\cite{cheshir1,cheshir2,cheshir3}. These proposals all involve
optical schemes.\ We give here instead a "Cheshire cat" example with
massive spin-1\ particles that can in principle be realized in
atomic interferometry experiments

Our scheme is based on a setup almost identical to the one presented in Fig.
\ref{fig j=1 setup}, with the initial and final states given by Eqs.
(\ref{ini}) and (\ref{30}) respectively. The focus here is on path $A$ on the
one hand and the recombined path $B+C$ on the other, as portrayed in
Fig.\ \ref{fig cheshire}. The results given in Sec. \ref{condA} hold, ie we
choose $\alpha$ and $\phi$ obeying Eqs. (\ref{condition1})-(\ref{fisol}) so
that $\left\langle \Pi_{\bar{A}}\right\rangle _{w}=0$. This means that the apparati  $\mathcal{D}_{2}$ and $\mathcal{D}_{3}$ interacting with the wavepacket
only detect the particle on path $A$ and detect nothing on path $B+C$.\ The
particle hence took path $A$.

Let us now introduce the spin projection $J_{\gamma}$ along the axis
$\mathbf{\hat{\gamma}}$ as the system property we wish to follow from $t=t_{i}
$ to $t=t_{f}$ as the system goes from the initial to the final states. We
choose $\gamma$ such that%
\begin{equation}
\left\langle m_{f}\right\vert J_{\gamma}\left\vert m_{\alpha}=1\right\rangle
=0\label{conditionCC}%
\end{equation}
giving $\gamma$ as a function of $\alpha$ and $\phi:$%
\begin{equation}
\gamma=-2\arctan\left(  \frac{\tan\frac{\alpha}{2}+\tan\frac{\phi}{2}%
-\sqrt{\sec^{2}\frac{\alpha}{2}\sec^{2}\frac{\phi}{2}}+}{\tan\frac{\alpha}%
{2}\tan\frac{\phi}{2}-1}\right)  +2\pi n
\end{equation}
(recall that $\alpha$ and $\phi$ are related by Eq. (\ref{fisol}) so that for
a fixed postselected state $\gamma$ only depends on $\alpha)$.

Let us place  an apparatus $\mathcal{C}_{0}$ just after the initial state has
been launched, before the SG type of device $\mathcal{D}_{1}$. The weak value
$\left\langle J_{\gamma}(t_{0})\right\rangle _{w}$ measured at $t_{0}\simeq
t_{i}$ is%
\begin{equation}
\left\langle J_{\gamma}(t_{0})\right\rangle _{w}=\frac{\left\langle
m_{f}\right\vert J_{\gamma}\left\vert m_{i}\right\rangle }{\left\langle
m_{f}\right\vert \left.  m_{i}\right\rangle }\label{60}%
\end{equation}
which is typically non-zero. Let us position an  apparatus $\mathcal{C}_{5}$
just before postselection takes place (see Fig.\ \ref{fig cheshire}).\ The
weak value at $t_{5}\simeq t_{f}$ is again given by%
\begin{equation}
\left\langle J_{\gamma}(t_{5})\right\rangle _{w}=\left\langle J_{\gamma}%
(t_{0})\right\rangle _{w}=\frac{\left\langle m_{f}\right\vert J_{\gamma
}\left\vert m_{i}\right\rangle }{\left\langle m_{f}\right\vert \left.
m_{i}\right\rangle }\label{59}%
\end{equation}
and therefore non-zero.

Let us now place an apparatus $\mathcal{C}_{2}$ localized on the recombined
$B+C$ branches (see Fig.\ \ref{fig cheshire}).\ When interacting with the
system\ $\mathcal{C}_{2}$ measures the weak value of $J_{\gamma}$ along that
path, denoted $\left\langle J_{\gamma}^{\bar{A}}\right\rangle _{w}$. The
system wavefunction is given by Eq. (\ref{13}) and $\mathcal{C}_{2}$ couples
to the last term only (the term describing the path $B+C$). The weak value
$\left\langle J_{\gamma}^{\bar{A}}\right\rangle _{w}\equiv\left\langle
J_{\gamma}(t_{2})\Pi_{\bar{A}}(t_{2})\right\rangle _{w}$ is given by
\begin{align}
\left\langle J_{\gamma}^{\bar{A}}\right\rangle _{w} &  =\frac{\left\langle
\xi(t_{f})\right\vert U(t_{f},t_{2})\left\vert \xi_{B+C}(t_{2})\right\rangle
\sum_{k=-1,0}\left\langle m_{f}\right\vert J_{\gamma}\left\vert m_{\alpha
}=k\right\rangle \left\langle m_{\alpha}=k\right\vert \left.  m_{i}%
\right\rangle }{\left\langle m_{f}\right\vert \left.  m_{i}\right\rangle
}\label{61}\\
&  =\frac{\left\langle m_{f}\right\vert J_{\gamma}\left\vert m_{i}%
\right\rangle }{\left\langle m_{f}\right\vert \left.  m_{i}\right\rangle
}\label{62}%
\end{align}
where we have used Eq. (\ref{conditionCC}) and assumed an ideal projection
$\Pi_{\bar{A}}(t_{2})\equiv\left\vert \xi_{B+C}(t_{2})\right\rangle
\left\langle \xi_{B+C}(t_{2})\right\vert $. Comparing Eqs. (\ref{60}),
(\ref{59}) and (\ref{62}), it looks as if $J_{\gamma}$ had traveled entirely
along the path $B+C$. This is confirmed by positioning an apparatus
$\mathcal{C}_{3}$ in order to measure $\left\langle J_{\gamma}\right\rangle
_{w}$along path $A$ at $t=t_{3}$. The weak value $\left\langle J_{\gamma}%
^{A}\right\rangle _{w}\equiv\left\langle J_{\gamma}(t_{3})\Pi_{A}%
(t_{3})\right\rangle _{w}$ is%
\begin{equation}
\left\langle J_{\gamma}^{A}\right\rangle _{w}=\frac{\left\langle
m_{f}\right\vert J_{\gamma}\left\vert m_{\alpha}=1\right\rangle \left\langle
m_{\alpha}=1\right\vert \left.  m_{i}\right\rangle }{\left\langle
m_{f}\right\vert \left.  m_{i}\right\rangle }=0\label{71}%
\end{equation}
which is seen to vanish because of the condition (\ref{conditionCC})
imposed on $\gamma$.

If we now collect the results we see that [Eqs. (\ref{condition1}) and
(\ref{18})]%
\begin{equation}
\left\langle \Pi_{A}\right\rangle _{w}=1,\text{ \ }\left\langle \Pi_{\bar{A}%
}\right\rangle _{w}=0\label{73}%
\end{equation}
that can be interpreted as meaning that, given the initial and final
states $\left\vert \psi_{i}\right\rangle $ and $\left\vert
\psi_{f}\right\rangle ,$ `\emph{`the particle has traveled along path $A,$
as it cannot be found along the
other path}''. We have also obtained%
\begin{equation}
\left\langle J_{\gamma}^{A}\right\rangle _{w}=0\text{,\ \ }\left\langle
J_{\gamma}^{\bar{A}}\right\rangle _{w}=\left\langle J_{\gamma}(t_{i}%
)\right\rangle _{w}=\left\langle J_{\gamma}(t_{f})\right\rangle _{w}\label{74}%
\end{equation}
that can be interpreted as meaning that ``\emph{the property
$J_{\gamma}$ of the particle traveled along route $B+C$, as it
cannot be obtained along path $A$}''. We have therefore realized a
setup for the manifestation of a ``Cheshire cat'' with spin-1\
systems since the grin ($J_{\gamma}$) appears to be ``disembodied''
from the cat (the particle).\

\section{Discussion and Conclusion\label{disc}}

We have given in this work a proposal for an implementation with
spin-1 particles of the three-box paradox and of a Cheshire cat type
of setup.\ Whether the three-box problem discussed in Sec.
\ref{3bspin} is really constitutive of a paradox, or perhaps more
strikingly, whether there is anything like ``disembodiment'' of a
property in the Cheshire cat setup presented above hinges on the
status of weak measurements. Indeed, WM has remained
controversial since their inception; the terms of the controversy
ultimately depend on the options taken on the meaning of the
theoretical entities of the quantum formalism.\ While the discussion
of these options with regard to the status of weak measurements is
out of the scope of the present paper \cite{forthcoming}, we will
nevertheless make a few remarks relative to the setups discussed
above.

The crucial difference between WMs and projective measurements is
that the latter suppresses the entangled linear superposition of
system-apparatus states (as if a collapse to a single term in the
pointer basis had taken place) while the former retains the full
wave aspect of the quantum system. For example in dynamical systems,
where the system wavefunction is characterized by a "sum over paths"
as prescribed by Feynman's propagator, an array of apparati
weakly interacting with the system should allow in principle to
detect the wavefunction simultaneously propagating along the
available paths \cite{weak traj} provided the paths are sufficiently
isolated from one another. The ``strength'' of weak measurements is
to capture this wave phenomenon -- too often thought of as being a
computational artifact -- with apparati weakly coupled to the
system. Here (see Fig. \ref{fig j=1 setup}) the  apparati
 placed at $\mathcal{D}_{2},$ $\mathcal{D}_{3}$ and $\mathcal{D}%
_{4}$ monitor the wave properties along the different paths.

The aspect of weak values as measuring the transition amplitudes --
generically written $\left\langle \psi_{f}\right\vert O\left\vert
\psi _{i}\right\rangle $ in the notation of Sec.\ \ref{swm} -- is
well illustrated in the Cheshire cat grin scheme.\ The weak value
(\ref{8}), whose squared norm can be seen as a renormalized
transition probability, vanishes if the
transition is forbidden.\ This is the case, according to Eqs. (\ref{73}%
)-(\ref{74}), of the transition generated by the measurement of $J_{\gamma}$
along path $A$ and of the transition generated by the position projection
along the path $B+C$. Associating a vanishing weak value with a forbidden
transition toward a final state is therefore perfectly cogent within standard
quantum mechanics -- provided it is kept in mind that the wavepackets take all
the available paths simultaneously.

Finally, a possible\ tentative manner to implement experimentally
the schemes developed above for spin-1\ particles would be to employ
atoms in setups based on well-established atomic interferometry
experiments \cite{atomexp}.\ For example based on the setup of Ref.
\cite{atomexp2} (a so called ``\emph{Stern-Gerlach atom
interferometer}''), hydrogen atoms can be prepared in the initial
state $2s_{1/2},J=1,m_{z}=0$ (where $J$ denotes the total angular
momentum of the hyperfine Hamiltonian) and passed in a region
containing a magnetic field.\ This yields a coherent superposition
of atoms in different states $\left\vert m_{\alpha}\right\rangle $
that is finally projected to a desired final state by using a
polarizer and a time of flight detection scheme. The weak
measurement of the projectors $\Pi_{A},$ $\Pi_{\bar{A}}$ etc.\ could
be realized by a selective laser excitation of a given $\left\vert
m_{\alpha }\right\rangle $ manifold. If the excitation pulse is an
$n$ photon coherent state, the measurement is weak provided $n$ is
large (the detection of the overlap between the original state
$\left\vert n\right\rangle $ and the $\left\vert n-1\right\rangle $
photon state after absorption of a photon gives almost no
information on the path) and the transition to an excited state does
not change the kinetic energy of the atomic wavepacket. We note that
a closely related problem (a three box quantum game in which
individual boxes can be addressed, though the resulting
wavefunctions cannot be recombined) has been very recently realized
experimentally with a three-level system using the $^{14}$N nuclear
spin ($I=1$) of the Nitrogen Vacancy center in diamond, the
preparation and readout being performed by manipulating the NV$^{-}$
electronic spin ($S=1$)\ with $m_{S}$ and $m_{I}$ selective
microwave pulses \cite{NV}.

As for the Cheshire cat property $\left\langle
J_{\gamma}^{A}\right\rangle _{w}=0$, it could be possible to observe
experimentally this property indirectly by inducing at
$\mathcal{C}_{3}$ (see Fig.\ 2) a weak rotation of $\mathbf{J}$
along the axis $\mathbf{\gamma}$ by a small angle
$\varepsilon_{\gamma}$. If the rotation is sufficiently weak so that
$\exp(-iJ_{\gamma}\varepsilon_{\gamma
})\approx1-iJ_{\gamma}\varepsilon_{\gamma}$ holds (while still being
detectable in the statistics of the post-selection), then a rotation
along path $A$ will not affect the state prior to post-selection,
since $\left\langle m_{f}\right\vert
\exp(-iJ_{\gamma}\varepsilon_{\gamma })\left\vert
m_{\alpha}=1\right\rangle \approx\left\langle m_{f}\right\vert
\left.  m_{\alpha}=1\right\rangle -i\varepsilon_{\gamma}\left\langle
m_{f}\right\vert J_{\gamma}\left\vert m_{\alpha}=1\right\rangle $,
the last term vanishing by Eq. (\ref{71}). Hence the rotation at
$\mathcal{C}_{3}$ will have no effect and will not modify the
post-selection statistics, a statement that is equivalent to having
a vanishing weak value $\left\langle J_{\gamma }^{A}\right\rangle
_{w}=0$. On the other hand, if the same weak rotation along $\gamma$
generated by $J_{\gamma}$ is performed at $\mathcal{C}_{0}$,
$\mathcal{C}_{2}$ or $\mathcal{C}_{5}$ then the post-selection
statistics determined experimentally will be affected, and will be
so in exactly the same way.\ Thus everything happens as if
$J_{\gamma}$ had travelled along the route
$\mathcal{C}_{0}-\mathcal{C}_{2}-\mathcal{C}_{5}$, but not along
path $A.$

Summarizing, we have given a proposal implementing the three-box paradox and a
Cheshire cat grin scheme for massive spin-1 systems. Besides giving a concrete
rendering of paradigmatic examples of weak measurements, employing a definite
physical system sheds light on the peculiar quantum properties unraveled by
weak measurements while avoiding the ambiguities of the original ideal
three-box setup that have given rise to several controversies. In principle
the proposed schemes could be implemented in atomic interferometry experiments.

\begin{acknowledgements}
We thank Jacques Robert (Universit\'{e} de Paris-Sud, Orsay) for fruitful
discussions concerning atomic interferometry experiments. AKP acknowledges the support from JSPS Postdoctoral Fellowship for Foreign Researcher and Grant-in-Aid for JSPS fellows no. 24-02320.
\end{acknowledgements}


\begin{thebibliography}{99}                                                                                               %
\bibitem {wheeler}J. A. Wheeler in J. A. Wheeler and W. H. Zurek (Eds.),
\emph{Quantum Theory and Measurement} (Princeton University Press, Princeton,
1984), p. 182.

\bibitem {tbp}Y. Aharonov and L. Vaidman, J. Phys. A 24, 2315 (1991).

\bibitem {cheshir1}Y. Aharonov, S.Popescu, and P.Skrzypczyk, arXiv:1202.0631 (2012).

\bibitem {cheshir2}Y. Guryanova, N. Brunner and S. Popescu, arXiv:1203.4215 (2012).

\bibitem {cheshir3}A. Di Lorenzo, arXiv:1205.3755 (2012).

\bibitem {peculiar}Y. Aharonov, S. Nussinov, S. Popescu and L. Vaidman Phys.
Rev. A 87, 014105, 2013

\bibitem {aharonov rohrlich book}Y. Aharonov and D. Rohrlich, \emph{Quantum
Paradoxes} (Wiley, Weinheim, Germany, 2005), Ch. 17.2.

\bibitem {WM}Y. Aharonov and L. Vaidman, Phys. Rev. A 41, 11 (1990).
References to recent work in the area can be found in the reviews Y. Shikano
arXiv:1110.5055 (2011), or A. G. Kofman, S. Ashhab and F. Nori arXiv:1109.6315 (2011).

\bibitem {exp}K. J. Resch, J.S. Lundeen and A. M. Steinberg, Phys. Lett. A,
324, 125 (2004).

\bibitem {atomexp}J. Baudon R. Mathevet and J. Robert J. Phys. B 32, R173 (1999).

\bibitem {classical3}K.\ A. Kirkpatrick, J.\ Phys.\ A 36 4891, 2003.

\bibitem {soko}D. Sokolovski, I. Puerto Gim\'{e}nez and R. Sala Mayato
Phys.\ Lett.\ A 372, 6578, 2008.

\bibitem {atomexp2}R. Mathevet, K. Brodsky, J. Baudon, R. Brouri, M. Boustimi,
B. Viaris de Lesegno and J. Robert Phys. Rev. A 58, 4039 (1998).



\bibitem {ravon vaidman07}T. Ravon and L. Vaidman, J. Phys. A 40, 2873 (2007).

\bibitem {abl}Y. Aharonov, P. G. Bergmann and J. L. Lebowitz, Phys. Rev. 134,
B 1410, (1964).

\bibitem {sudarshan}I. M. Duck, P. M. Stevenson and E.C. G. Sudarshan Phys.
Rev. D 40 2112(1989).

\bibitem {pan matzkin12}A. K. Pan and A. Matzkin, Phys. Rev. A 85, 022122 (2012).

\bibitem {tollaksen07}J. Tollaksen, J. Phys. A 40, 9033 (2007).

\bibitem {forthcoming}The topics concerning the interpretation of WM will be
discussed in a forthcoming paper in relation to other works in that area.

\bibitem {weak traj}A. Matzkin, Phys. Rev. Lett. 109, 150407 (2012).

\bibitem {NV}R.\ E.\ George, L.\ M\ Robledo, O.\ J.\ E. Maroney, M.\ S.\ Blok,
H.\ Bernien, M. L.\ Markham, D.\ Twitchen, J.\ J. L.\ Morton, A.\ D. Brigg,
and R.\ Hanson,\emph{\ }PNAS 110 3777 (2013).
\end{thebibliography}
\end{document}